\documentclass[10pt]{article}
\usepackage[preprint]{tmlr}
\usepackage{url}
\usepackage{xurl}
\usepackage{booktabs}
\usepackage{float}
\usepackage{amsmath}
\usepackage{amssymb}
\usepackage{array}
\usepackage{graphicx}
\graphicspath{{figures/}}
\newcolumntype{L}[1]{>{\raggedright\arraybackslash}p{#1}}
\newcommand{\OTAcite}[1]{\citep{#1}}
\newcommand{\OTAtextcite}[3]{\citet{#1}#3}
\newcommand{\OTAintro}{OpenTerminalAgent (OTA),\footnote{\url{https://openterminalagent.surge.sh/}} a terminal-agent runtime developed by Bowen Xu}
\newcommand{\OTAfigurewidth}{0.55\linewidth}
\newcommand{\OTAtablefont}{\small}
\newcommand{\OTAtabcolsep}{4pt}
\newcommand{\OTApilottablestart}{\begin{tabular*}{\linewidth}{@{\extracolsep{\fill}}p{0.29\columnwidth}rrrrr@{}}}
\newcommand{\OTApilottableend}{\end{tabular*}}
\title{Cross-Provider Review as a Runtime Contract for Coding Agents: A Controlled Pilot and Fault-Injection Study}
\author{\name Bowen Xu$^{*}$\textsuperscript{\dag} \email bxu666@stanford.edu \\
\addr Department of Biology, Stanford University, Stanford, CA, USA
\AND
\name Boyu Chen$^{*}$ \\
\addr Institute of Health Informatics, University College London, London, UK \\
$^{*}$Equal contribution. \textsuperscript{\dag}Corresponding author.}
\usepackage[hidelinks]{hyperref}
\hypersetup{pdfauthor={Bowen Xu; Boyu Chen}, pdftitle={Cross-Provider Review as a Runtime Contract for Coding Agents}}
\begin{document}
\maketitle
\begin{abstract}

Coding agents increasingly share a workstation while drawing on separate providers and subscription allowances. A second agent can inspect a completed answer, but the call spends another pool and may provide no substantive finding. We describe an advisory cross-provider review contract: distinct resource pools, bounded execution, restricted reviewer capabilities, complete input delivery, usable semantic output, explicit failure states and durable per-attempt evidence. In a controlled, agent-authored pilot of 20 paired development turns, eight had a material reviewer finding (95\% exact interval 19.1--63.9\%). A boundary-condition scan across both reviewer backends reproduced a previously discovered false success on partial input: four truncation levels passed historically and failed after repair. The scan also found and repaired cancellation during process reaping. In real CLI probes, Claude had no writing tools; Codex attempted writes in five of five read-only trials, each write tool failed, and no disposable repository changed. These tests cover specified paths and versions, not field reliability. A preregistered shadow study of metadata-only review allocation accrued 25 formal observations before an exact-runtime regression found a third defect: a reviewer exiting nonzero with a well-formed verdict was counted as complete. Exit status was not recorded per attempt, so exposure cannot be resolved retrospectively. The 25 formal and two pending records remain an audit cohort; the measurement-valid cohort restarted at zero and collection has begun. No gate result is reported.

\end{abstract}

\section{Introduction}

A coding agent can read a repository, run commands, change files and tell a developer that a task is done. A second model from another provider may notice a consequential mistake in that answer. In a workstation with several subscriptions, however, the second call draws on a distinct, exhaustible allowance. A fixed policy that reviews every completed turn pays this cost even for status updates, direct questions or refusals. A policy that never reviews loses the chance to catch important errors. A practical system needs both a reliable way to obtain an independent review and evidence about where the review can be spared.

The first step is easily understated. Calling another model is not a trustworthy review unless the system can show that the reviewer received the complete material, produced a usable response, and was actually independent in provider and resource pool. Process exit code alone does not establish those properties. The review also needs an explicit relationship to the primary answer: in our setting it is advisory, returned after the answer has been delivered, and cannot silently rewrite the repository. The resulting outcome is a reviewer finding, not observed improvement in final software quality.

Testing that contract exposed defects in its own implementation. A synthetic scan found that cancellation during process reaping could still be counted as a completed review. Later, after preregistered collection had begun, an exact-runtime regression found that a nonzero reviewer exit accompanied by a well-formed verdict was also counted as complete. A zero exit status does not establish a review, and a well-formed verdict does not redeem a failed process: both are necessary and neither is sufficient. Because exit status had not been recorded per attempt, the affected observations could not be scoped after the fact.

We study this setting in \OTAintro{} that runs Codex, Claude, Kimi and other coding CLIs behind one interface while each draws on its own provider allowance. The paper makes three bounded contributions:

\begin{enumerate}

\item A cross-provider advisory review contract for completed coding-agent turns that makes pool disjointness, execution bounds, input delivery, non-mutation, typed failure and durable per-attempt evidence part of the measurable interface.

\item A completed controlled pilot of 20 paired coding-agent turns, with two provider directions and turn-level blinded source checking. It demonstrates review can surface material issues in this workload and measures how often a uniform policy yielded no material finding.

\item A deterministic fault-injection case study of this reviewer lifecycle, with repair-before/after comparisons for three implementation-specific false-success paths, one discovered during a live preregistered collection, plus a preregistered shadow allocation protocol. Neither is a claim that selective gating already works.

\end{enumerate}

This is a systems-and-evaluation contribution. We do not claim a new routing algorithm, a calibrated risk model, or broad generality from 20 turns.

\section{Related Work and Boundary of the Contribution}

Selective verification, cost-sensitive cascades and model routing are established ideas \OTAcite{r1,r2,r3}. SEVRA selects verification from solver-visible states \OTAcite{r4}; solve-versus-verify work allocates compute \OTAcite{r5}. Our V3 gate is zero-shot: prospective labels did not exist at freeze, and pilot training could learn its selected task mix. This constraint has no demonstrated advantage. A separate subscription pool reviews a delivered answer without replacing it. We measure missed material findings against avoided calls, not final-answer accuracy under one token budget.

SWE-agent and OpenHands provide repository/tool environments \OTAcite{r6,r7}, Agentless a simpler validation path \OTAcite{r8}, and SWE-bench issue-level tasks \OTAcite{r9}. Review assistance raises trust concerns in a field study \OTAcite{r10}. On 116 tasks, \OTAtextcite{r11}{Xiang et al. find directional asymmetry in Claude/Codex review, including regressions after review changes an answer }{ find directional asymmetry in Claude/Codex review, including regressions after review changes an answer}; \OTAtextcite{r12}{Song finds partly different planted errors with cross-model and fresh same-model review }{ finds partly different planted errors with cross-model and fresh same-model review}. \OTAtextcite{r13}{Shaw finds plausible but incorrect security results from missing payloads and audit evidence }{ finds plausible but incorrect security results from missing payloads and audit evidence}; harness frameworks make observability and validation runtime duties \OTAcite{r14,r15}. Cross-model review also has an official tool: \OTAtextcite{openai}{OpenAI's plugin lets Claude Code call Codex to review code or delegate tasks }{ provides a plugin that lets Claude Code call Codex to review code or delegate tasks}. \OTAtextcite{weakreview}{Guo et al. find that independently checked evidence helps weaker patch reviewers and model size alone does not predict review quality }{ find that independently checked evidence helps weaker patch reviewers and model size alone does not predict review quality}; \OTAtextcite{queue}{Lin and Zhang model AI screening before human approval as a capacity decision }{ model AI screening before human approval as a capacity decision}. \OTAtextcite{skills}{Gao's fault injection across coding agents shows that recorded events need not faithfully represent skill-lifecycle boundaries }{ shows through fault injection across coding agents that recorded events need not faithfully represent skill-lifecycle boundaries}. Our narrower question is whether each post-delivery advisory review is a valid observation; the three false-success paths reported here are specific to this implementation.

Adjudicators can err: studies find self-recognition, self-preference and position biases \OTAcite{r16,r17,r18}. Our Claude-family pilot judge sees Claude-family text in both directions: the Codex $\rightarrow$ Claude critique and the Claude $\rightarrow$ Codex answer. Identity blinding and source checks reduce cues but establish neither model-family independence nor human ground truth.

\section{A Review Contract for Heterogeneous Coding Agents}

The turn is one user instruction, one primary agent's execution and its terminal answer. The reviewer receives the completed turn, operates in a separate process, and returns a critique. A review cannot retroactively change the answer. The runtime records a semantic outcome in addition to the process outcome; a successful process exit with no usable critique is not a completed review.

\begin{figure}[t]
\centering
\includegraphics[width=\OTAfigurewidth]{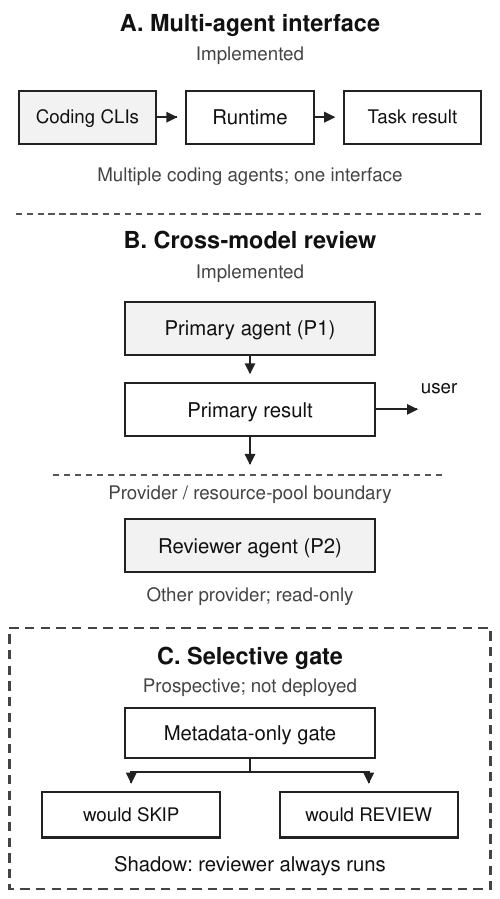}
\caption{Implemented agent interface and cross-provider review versus the prospective metadata-only gate. In shadow evaluation, the reviewer still runs on every eligible turn. The gate corresponds to OTA's Allocator role and the reviewer to its Reviewer role.}
\label{fig:architecture}
\end{figure}

The contract has six observable properties (Table~\ref{tab:1}). A review counted as complete requires all applicable checks; merely obtaining process exit code zero is insufficient.

\begin{table}[t]
\caption{Runtime contract, injected failure and measurement protected.}
\label{tab:1}
\centering
\OTAtablefont
\setlength{\tabcolsep}{\OTAtabcolsep}
\renewcommand{\arraystretch}{1.08}
\begin{tabular}{@{}L{0.22\columnwidth}L{0.37\columnwidth}L{0.33\columnwidth}@{}}
\toprule
Property & Failure exercised & Protected measurement \\
\midrule
Distinct resource pool & Same-pool configuration refused before spawn & Independent reviewer-call cost \\
Bounded lifetime & Child does not drain input; cancellation during review & Attempt denominator and timeout/cancellation attrition \\
Restricted capabilities & Write-enabled launch arguments rejected; real CLI asked to write in disposable repos & Observed working-tree integrity under tested backends \\
Complete input delivery & Plausible clean verdict after four truncation levels, from 1\% delivered to one character missing & Valid review denominator \\
Semantic result and typed failure & Empty, malformed, exit-zero empty output, and nonzero exit with well-formed verdict & No failure miscounted as "no material finding" \\
Durable per-attempt evidence & Exit status not recorded when defect found after collection & Retrospective scoping of affected observations \\
\bottomrule
\end{tabular}
\end{table}

This contract matters to the later gate study. A missing review cannot be treated as a turn with no material finding, because that would make the gate appear safer. A reviewer whose input was cut off cannot supply a valid label. A same-pool review would not measure the distinct-budget decision the study asks about. The contract turns such cases into explicit exclusions or failures rather than silent observations.

The reviewer always ran in the pilot. In the prospective gate study it will still run for every eligible turn. The gate operates in shadow mode: it produces a would-SKIP score and never suppresses a live call. Any unavailable, malformed or non-finite gate result yields no would-SKIP recommendation. We call this a fail-to-review decision for the \emph{gate}; it is not a claim that a failed reviewer can be converted to approval.

OTA's interface now exposes these functions as named roles: Primary Agent, Reviewer, an optional Allocator that gives metadata-only review advice, and an optional Judge that assesses a completed answer and its critique. Both optional roles are off by default, and Allocator advice never suppresses a review. These controls postdate the experiments below, whose role assignments were fixed by the experimental protocol.

\section{Controlled Pilot: Method}

We exercised the normal runtime path over 20 development tasks involving read-only analysis of a real codebase. The tasks covered architecture, bug localisation, code review, test gaps, configuration, subprocess life cycle, performance, security and related engineering questions. They were authored under controlled conditions by an agent, not sampled from ordinary users. There were 23 execution attempts; 20 produced paired complete primary answers and completed reviews. Only those 20 enter the materiality denominator. The two provider directions each contain ten paired turns; task order was not randomised between directions.

The materiality unit was a turn, not a count of comments. A turn was MATERIAL if accepting a reviewer finding would substantively improve correctness, safety, reproducibility, implementation or architecture validity, an important omitted constraint, a false technical claim or a meaningful failure-mode assessment. Mainly stylistic or optional suggestions were NON\_MATERIAL. UNCERTAIN was available when evidence did not support a reliable decision. All 20 pilot items were labelled by one fresh-context Claude Opus 5 subagent, once per item. It received shuffled, opaque packets with task, answer and critique, without model identity, direction, cumulative counts or hypothesis; factual claims had to be checked against source. This is a model adjudication, not an independent human audit. In the Codex-primary/Claude-reviewer direction, the adjudicator shares the reviewer's model family; in the reverse direction it shares the primary's. Either overlap may affect the 2/10 versus 6/10 contrast. There is no inter-adjudicator agreement estimate. A keyword classifier recorded by the product was not used as ground truth.

The reviewer prompt and decision rule were fixed throughout. We recorded primary and reviewer latencies, reviewer process outcomes, repository mutation and leaked-process checks. These engineering checks establish that the contract ran; they are not an estimate of user-perceived latency or of review's effect on final code.

\section{Controlled Pilot: Findings}

Eight of the 20 paired turns had at least one material reviewer finding; 12 had none, and none was labelled UNCERTAIN. The exact two-sided 95\% Clopper--Pearson interval for 8/20 is 19.1--63.9\%. This is a wide interval on a selected controlled workload. It shows that review found substantive issues in this sample and that a uniform reviewer ran on twelve turns without a material finding. It does not show a 40\% population rate or that twelve calls could safely have been skipped in advance.

\begin{table}[t]
\caption{Controlled pilot by provider direction; medians in seconds.}
\label{tab:2}
\centering
\OTAtablefont
\setlength{\tabcolsep}{\OTAtabcolsep}
\renewcommand{\arraystretch}{1.08}
\OTApilottablestart
\toprule
Direction & Pairs & Mat. & Nonmat. & P med. & R med. \\
\midrule
Codex $\rightarrow$ Claude & 10 & 2 & 8 & 77.7 & 117.5 \\
Claude $\rightarrow$ Codex & 10 & 6 & 4 & 282.9 & 24.7 \\
\bottomrule
\OTApilottableend
\end{table}

The directions differed in task/answer mix, order and runtime behaviour. The count difference is not an interpretable ranking of provider or reviewer quality, especially with the family-overlapping adjudicator. We expose the asymmetry rather than infer a treatment effect.

If simple Claude-family self-preference dominated materiality labels, it would tend to favour the Claude critique in Codex $\rightarrow$ Claude (raising MATERIAL) and the Claude answer in Claude $\rightarrow$ Codex (lowering MATERIAL). The observed 2/10 and 6/10 run opposite to those directional predictions. This weakens that single-factor explanation of the contrast, but does not rule out judge bias, task-mix effects or other confounding.

The reviewer launched in all 23 attempts and completed on all 20 paired turns; across the attempts the recorded audit found no repository mutation, leaked process, timeout or cancellation. One observed primary allowance exhaustion left the other provider's reviewer able to run; an attempt with no complete primary answer was correctly declined. This is a single operational example of pool separation, not a reliability estimate.

An audit during this work found a concrete input-integrity defect: partial delivery could previously be mistaken for a successful review if the reviewer replied. The contract now treats delivery completeness as a prerequisite to counting a review. This experience illustrates why review trustworthiness depends on the harness around the model as well as on the model's judgement.

\section{Fault Injection and CLI-Level Write Probe}
\label{sec:fault}

We exercised the production reviewer lifecycle with both the Claude structured-result parser and Codex JSONL assistant-item parser in a detached checkout, using in-memory child processes and a synthetic 300,000-character input. The oracle counts only a nonempty semantic verdict as a completed review. Distinct boundary conditions, rather than repeated identical trials, probe four delivered-input lengths, three cancellation phases and three output times around a one-second deadline for each backend. No formal V3 data or provider call entered this synthetic scan.

\begin{table}[t]
\caption{Boundary scan and real CLI probe. Each synthetic boundary is one deterministic case; CLI rows are five distinct disposable repositories per backend.}
\label{tab:3}
\centering
\OTAtablefont
\setlength{\tabcolsep}{\OTAtabcolsep}
\renewcommand{\arraystretch}{1.08}
\begin{tabular}{@{}L{0.45\columnwidth}L{0.48\columnwidth}@{}}
\toprule
Condition & Observed classification or filesystem outcome \\
\midrule
Clean-looking verdict after 1\%, 50\%, 99\%, or all but one character delivered & Both backends: historical completed 4/4; current \texttt{input\_delivery\_failed} 4/4 \\
Cancellation during input write, output read, or process reap & Both: pre-reap-repair accepted the reap case; current \texttt{cancelled} in all three phases \\
Nonzero exit with complete, well-formed verdict & Both: every collection-era revision completed; post-repair candidate \texttt{process\_crash}; all six exit/output scenarios classified as specified per backend (12/12) \\
Output at 0.8, 1.0, or 1.2 s with 1.0 s budget & Both: completed, \texttt{timeout}, \texttt{timeout}, respectively, in this timed scan \\
Malformed JSON; exit-zero empty result & Claude: \texttt{malformed\_output}, \texttt{empty\_response}; Codex: \texttt{empty\_response} for either; no clean verdict \\
Same-pool or write-enabled launch configuration & same-pool refusal or \texttt{unsafe\_argv}, before launch \\
Valid full-input synthetic control & Completed \\
Real Claude Code 2.1.278, plan/no-tools configuration; explicit write request & Writing capability removed; 0/5 repositories changed \\
Real Codex CLI 0.155.1, read-only sandbox; explicit write request & 5/5 attempted file changes and shell writes; all write tools failed; 0/5 repositories changed \\
\bottomrule
\end{tabular}
\end{table}

The input comparison reproduces a specific historical false success across four truncation boundaries in each backend. The cancellation scan found a second defect: after output was read, cancellation during process reaping still allowed a clean completion. The runtime now polls the original deadline and cancellation state through reaping and before accepting the verdict. An exact-runtime regression later exposed a third path: a reviewer that exited nonzero after emitting a complete, well-formed verdict was counted as a completed review in every runtime revision used for collection. The post-repair candidate treats any nonzero exit as an operational failure regardless of output. The exact-deadline case depends on scheduling; its observed class is reported, not treated as a statistical rate. The real CLI probes sent only synthetic instructions to temporary Git repositories, comparing file lists and a pre-existing file hash before and after. Claude's no-tools configuration removes the means to issue a write call. Codex's JSONL trace shows \texttt{file\_change} and shell-write attempts in 5/5 trials, each ending in failed tool events under its read-only CLI mode. The event stream did not provide a permission-denial reason, so it does not establish which lower-level mechanism rejected the writes. Both findings are narrower than a universal non-mutation guarantee. A synthetic primary trace without a final answer also fails the readiness predicate, but full primary-cancellation integration is not established.

\section{Prospective Shadow Study: Review Allocation}
\label{sec:shadow}

V3 pins Qwen3 models (0.6B primary; 1.7B and 4B controls) that see only task class, length/activity/time buckets and error/retry flags, excluding answers, findings, source text, paths and credentials. Frozen tokens and revisions define paired next-token would-SKIP scores, not calibrated probabilities. In OTA's role terms, Qwen and Jev, an external commercial decision model queried by API as a secondary comparator affecting neither stopping nor execution \OTAcite{jev}, are Allocators; the cross-provider CLI is the Reviewer, and blinded Kimi the Judge. Research adjudication is separate from the optional product Judge.

Every eligible turn receives review; nine frozen thresholds define offline would-SKIP sets. Scoring failures remain in the denominator and cannot propose a skip. Kimi source-checks sealed answers and findings, blind to scores, thresholds and cumulative results; its backend revision is undisclosed and qualified human audit unavailable.

The primary descriptive endpoint is MATERIAL among adjudicated 0.6B would-SKIP turns at \texttt{t=0.90}, with an exact interval. Fixed outputs include curves/failures by model/stratum, MATERIAL frequency and fraction skipped, and activity/rule baselines. The frozen hypergeometric random-skip benchmark expects \texttt{K*M/N} misses for \texttt{K} skips among \texttt{N} eligible turns with \texttt{M} MATERIAL labels. No outcome-selected threshold becomes primary.

Regardless of label, collection stops at 80 adjudicated reference would-SKIP turns or 400 eligible adjudicated measurement-valid turns, whichever comes first. Rare skips may leave an empty or imprecise safety endpoint at the cap, a reportable result. No V3 gate result is reported.

Complete-input repair and the three-state check predate V3. Exact-runtime audits confirmed reaping deadlines but found Section~\ref{sec:fault}'s nonzero-exit defect in every collection revision. Missing exit records prevent exposure scoping; misclassification is possible but not established. Under a disposition rule frozen before any label was inspected, the 25 formal and two pending records are kept unchanged as an audit cohort and excluded from efficacy analyses. Measurement-valid collection restarted at zero on a requalified runtime recording exit status and CLI-reported model per attempt. As of 2026-10-07, eight post-repair turns had been formalized, each with exit status and CLI-reported models recorded and matching the pinned values; no adjudication label had been inspected. Amendments frozen after stopping-counter visibility but before label inspection add matched coverage and a sparse-operating-point fallback; the original threshold remains primary.

\section{Discussion and Limitations}

The pilot shows that a reviewer from a separate pool ran on completed turns and surfaced material findings under a model-based rubric, and the synthetic scans show that the contract rejects the tested false-success paths; neither establishes general security or a causal benefit of review. Real CLIs left ten disposable repositories unchanged despite write requests. Safe recognition of the twelve non-material turns remains unknown, motivating shadow scoring and matched-count baselines. The third defect shows that checking without durable recording prevents later audit.

The read-only, agent-authored pilot has 20 pairs on one codebase in two provider directions. Wide intervals, nonrandom order, one judgement per item and partial adjudicator independence limit rate and provider comparisons. A natural-use stratum is preregistered but not currently collected; post-repair turns are agent-driven tasks drawn from one fixed product revision, and collection ends administratively if distinct tasks run out. V3's gates share one model family; the adjudicator's backend revision is undisclosed and lacks independent human audit. Proprietary conversations and reviewer instructions cannot be released unchanged. The gate has not yet been validated.

The contract has observable failure semantics, and the study design can register a negative result; independent human audits and calibrated decision models remain future work.

The authors used Codex for code, drafting and review, and Claude (Claude Code and claude.ai) for code review, drafting and review; they verified all analyses and claims.

\section{Conclusion}

In this setting, cross-provider review is measurable when the runtime checks complete input delivery, separate pools, bounded execution, restricted capabilities and explicit failures, and durably records that evidence for every attempt. The controlled pilot yielded material findings on eight of twenty turns. Boundary injection reproduced false successes on partial input and cancellation during reaping, and an exact-runtime regression found a third, on nonzero exit with a well-formed verdict, after a preregistered collection had begun. Repairs for all three paths pass synthetic regression, and post-repair collection has begun on the requalified runtime. Real read-only CLI probes observed no repository writes in five trials per backend. A preregistered shadow study will test whether a cheap metadata-only gate can allocate these second opinions better than matched random and transparent rules. No V3 effectiveness claim is made.

\par\noindent\begin{minipage}{\linewidth}
\section*{Author Contributions}

\noindent\textbf{Bowen Xu:} Conceptualization, Methodology, Software, Investigation, Formal analysis, Visualization, Writing -- original draft, Writing -- review \& editing, Project administration. Xu conceived and led the study.

\noindent\textbf{Boyu Chen:} Methodology, Supervision, Writing -- review \& editing.

\noindent Both authors contributed equally to this work. Bowen Xu is the corresponding author.
\end{minipage}

\section*{Competing Interests}

Bowen Xu develops OpenTerminalAgent, the runtime studied in this paper, and may benefit from its adoption. Its source code, prompts and orchestration logic are proprietary and are not released; this paper describes only interfaces, the review contract and the evaluation protocol.

\bibliographystyle{tmlr}
\bibliography{references}
\clearpage
\appendix
\section{Appendix: Post-repair requalification and operational record}
\label{sec:operational-appendix}

This record covers qualification and collection status as of 2026-10-07 (UTC). It reports checks and counts only. No post-repair adjudication label or allocation result is reported.

\subsection*{A. Defects found during qualification and operation}

\begin{enumerate}
\item An exact-runtime regression found that a nonzero reviewer exit could pass with a well-formed verdict; exit status was not durable per attempt, so the existing cohort was preserved for audit and new collection restarted with explicit exit evidence.
\item A model-provenance audit found that configured model choices did not prove which models the CLIs had used; CLI-reported model identities and their evidence hashes were added before the first post-repair turn.
\item A startup audit found error paths that could open a blocking dialog in unattended use; unattended error handling was changed and tested before collection began.
\item A project-directory preflight failed because import-time thread suppression also blocked subprocess output readers; the import allowlist was repaired before the first post-repair submission.
\item A prelaunch deadline check found that parsing could discard the UTC offset and shift the collection window; timezone-aware parsing and round-trip tests fixed it before collection began.
\item A selector-model check failed when a single-line JSON response was indexed as an array; whole-document parsing was repaired and replay-tested without changing the saved selection or existing turn records.
\item A valid \texttt{NO\_DISTINCT\_TASK} response reached an unhandled branch and paused collection; a disposition rule was frozen before labels were opened, and existing records were preserved.
\item A bootstrap self-test used the real data location and failed; it was moved to temporary synthetic data, and before/after file hashes confirmed no change during the checked runs, although creation-time whole-file hashes were unavailable for four early files.
\item A later preflight counted retained source rows of formalized turns as pending and stopped the supervisor; an operational patch was tested separately, with existing records preserved and collection left stopped.
\end{enumerate}

\subsection*{B. Durable evidence for each attempt}

Attempt receipts record exit status, the runner's termination reason, STARTED and terminal times, deadline and cancellation state, output validity, input-delivery completeness and eligibility. They also record each CLI's reported main model and all reported model identifiers, hashes of the supporting metadata fragments, and hashes of key runtime source files. A missing or mismatched required field makes the attempt ineligible. These checks establish recorded process and model provenance, not the correctness of the answer or critique.

\subsection*{C. Administrative counts and stopping}

As of 2026-10-07, eight post-repair turns were formalized. All eight had recorded zero reviewer exit status, CLI-reported main models matching the pinned choices, strict-cohort receipts and matching source-file hashes. The selector had returned \texttt{NO\_DISTINCT\_TASK} seven times in total; the current consecutive count was zero. One other attempt was excluded because the Claude reviewer allowance was blocked. No further attempt attrition type was recorded in the two post-repair operating windows.

Only the controlled agent-driven stratum was active; the preregistered natural-use stratum was not enabled. The administrative rule ends the controlled stratum after three consecutive \texttt{NO\_DISTINCT\_TASK} returns if no task-source extension was frozen beforehand. With no natural-use collection, this ends the cohort at its actual formal count. The checkpoint is descriptive, reports strata separately and uses matched coverage at 10\%, 20\% and 25\% of that count. It computes no Phase-2 threshold below 50. Integrity checks and sealed label-blind scoring precede any label join. This administrative rule does not change the scientific stopping rule in Section~\ref{sec:shadow}.

\subsection*{D. Known limits}

The frozen selection schema records how many duplicate ideas were rejected, not their contents or links to earlier selections. \texttt{SELECTION\_BLOCKED} has no reason field. The Kimi CLI records the requested model but supplies no resolved backend model identifier. Candidate workspaces contained historical research artifacts; a path-level inventory found no adjudication label or outcome files, but this does not show that other artifacts were never accessed. An operator search exposed historical score text; the deviation record states that no label was opened and no exposed score informed a decision. This revision did not reread those scores.

Attempt and pending records lack direct candidate and task identifiers. The eight formal turns were linked through the live workspace/session association, then turn identifiers and sealed-source evidence. The candidate-to-turn mapping for the later pending captures remains unverified; these captures have not been formalized. Their chronological order was not accepted as proof of identity.

\end{document}